\newcommand\IPA{IPA-CuCl$_3$}
\documentclass[twocolumn,prb,showpacs,preprintnumbers,amsmath,amssymb]{revtex4}

\usepackage{graphicx}
\usepackage{dcolumn}
\usepackage{bm}

\begin{document}

\title{Dynamics of quantum spin liquid and spin solid phases in \IPA\ under field.}

\author{A. Zheludev}

\affiliation{Neutron Scattering Sciences Division, Oak Ridge
National Laboratory, Oak Ridge, Tennessee 37831-6393, USA.}

\author{V. O. Garlea}
\affiliation{Neutron Scattering Sciences Division, Oak Ridge
National Laboratory, Oak Ridge, Tennessee 37831-6393, USA.}

\author{T. Masuda}
\affiliation{International Graduate School of Arts and Sciences,
Yokohama City University, 22-2, Seto, Kanazawa-ku, Yokohama City,
Kanagawa, 236-0027, Japan.}

\author{H. Manaka}
\affiliation{Graduate School of Science and Engineering, Kagoshima
University, Korimoto, Kagoshima 890-0065, Japan.}

\author{L.-P.~Regnault}
\author{E.~Ressouche}
\author{B.~Grenier}
\affiliation{CEA-Grenoble, DRFMC-SPSMS-MDN, 17 rue des Martyrs,
38054 Grenoble Cedex 9, France.}

\author{J.-H. Chung}
\author{Y. Qiu}
\affiliation{NCNR, National Institute of Standards and Technology,
Gaithersburg, Maryland 20899, USA.} \affiliation{Department of
Materials Science and Engineering, University of Maryland, College
Park, Maryland, 20742, USA}

\author{K. Habicht}
\author{K. Kiefer}
\affiliation{BENSC, Hahn-Meitner Institut, D-14109 Berlin,
Germany.}

\author{M. Boehm}
\affiliation{Institut Laue Langevin, 6 rue J. Horowitz, 38042
Grenoble Cedex 9, France.}

\date{\today}

\begin{abstract}
Inelastic and elastic neutron scattering is used to study spin
correlations in the quasi-one dimensional quantum antiferromagnet
\IPA\ in strong applied magnetic fields. A condensation of magnons
and commensurate transverse long-range ordering is observe at a
critical filed $H_c=9.5$~T. The field dependencies of the energies
and polarizations of all magnon branches are investigated both
below and above the transition point. Their dispersion is measured
across the entire 1D Brillouin zone in magnetic fields up to 14~T.
The critical wave vector of magnon spectrum truncation [Masuda
{\it et al.}, Phys. Rev. Lett. 96, 047210 (2006)] is found to
shift from $h_c\approx 0.35$ at $H<H_c$ to $h_c=0.25$ for $H>H_c$.
A drastic reduction of magnon bandwidths in the ordered phase
[Garlea {\it et al.}, Phys. Rev. Lett. 98, 167202 (2007)] is
observed and studied in detail. New features of the spectrum,
presumably related to this bandwidth collapse, are observed just
above the transition field.

\end{abstract}


\maketitle

\section{Introduction}
Recent years were marked by substantial progress in the study of
gapped quantum-disordered antiferromagnets (AFs) in external
magnetic fields. The most prominent common feature is the
destruction of the non-magnetic spin liquid ground state and the
emergence of a long-range AF ordered phase at some critical field
$H_c$. Such quantum phase transitions can often be viewed as a
condensation of magnons. The Zeeman effect drives the energy of a
particular gapped magnon to zero at the AF wave vector, whereupon
they are macroscopically incorporated in the ground state. By now,
the phenomenon has been observed and studied experimentally in
numerous prototype materials. In spin systems with relevant
magnetic anisotropy, such the $S=1$ Haldane spin chain compound
NDMAP,\cite{Honda97,Zheludev2003,Zheludev2004} the model
bond-alternating $S=1$ chain NTENP,\cite{Narumi2001,Hagiwara2005}
or the $S=3/2$ spin-dimer system
Cs$_3$Cr$_2$Br$_9$,\cite{Grenier2004,Ziman2005} the transition is
of an Ising universality class,\cite{Affleck91} and the high-field
ordered phase is gapped. More recent work focused on materials
described by Heisenberg or XY (axially symmetric) Hamiltonians:
3D-interacting $S=1/2$ dimers in TlCuCl$_3$,\cite{Ruegg2003} the
$S=1/2$ quasi-2D network PHCC,\cite{Stone2006}, the $S=1/2$ square
lattice bilayer BaCu$_2$Si$_2$O$7$,\cite{Sebastian2005} and the
$S=1$ material NiCl$_2$-4SC(NH$_2$)$_2$.\cite{Zapf2006} Here the
transition can be viewed as a Bose-Einstein Condensation (BEC) of
magnons,\cite{Giamarchi1999} despite the ongoing controversy
regarding such a nomenclature.\cite{Mills2007} Correspondingly,
the high field ordered phase remains gapless.

In all the diverse cases, due to the soft-mode nature of the phase
transition, the key to the underlying physics is in understanding
the spin excitations on either side of the phase boundary. Until
recently, of the more isotropic materials, only in TlCuCl$_3$ has
the spectrum been studied in sufficient detail. Unfortunately,
recent data indicates that the transition in this compound is
actually not BEC, and that the high field spectrum is gapped, due
to Dzyaloshinski-Moriya type anisotropy\cite{Sirker2005} and
magnetoelastic effects.\cite{Johannsen2005} In a recent short
paper we reported neutron scattering studied of magnetic
excitations in the $S=1/2$ spin ladder \IPA, where the transition
is an almost exact realization the BEC case.\cite{Garlea2007} In
fact, any long-range properties, such the emergence of a gapless
collective Goldstone mode beyond the transition point, appear to
be fully consistent with original model of Giamarchi and
Tsvelik.\cite{Giamarchi1999} Interestingly, the transition was
also found to dramatically affect short-range spin correlations
and excitations near the magnetic zone boundary. This phenomenon
was attributed to AF long-range order violating discrete
translational symmetry. Such a spontaneous symmetry breaking is
absent in either the conventional BEC, or in magnon condensation
transitions in NTENP, TlCuCl$_3$ and many other spin gap systems.

The present paper is a report on extensive high-field neutron
scattering studies of \IPA. In addition to providing more details
on the previous experiments and data analysis procedures, we focus
on a number of new results. Among these are a study of
polarizations of magnetic excitations, observation of certain
unusual features of the spectrum just beyond the transition point,
a study of the effect of long-range ordering on single-magnon
spectrum termination,\cite{Masuda2006} and a measurement of magnon
dispersion at the highest attainable field of 14.5~T.

\IPA\ crystallizes in a triclinic space group $P\bar{1}$ with $a =
7.766$ \AA , $b = 9.705$ \AA , $c = 6.083$ \AA , $\alpha =
97.62^{\circ}$, $\beta = 101.05^{\circ}$, and $\gamma =
67.28^{\circ}$.\cite{Manaka97} As discussed in detail in
Ref.~\onlinecite{Masuda2006},  the key features of the structure
are $S=1/2$ ladders of Cu$^{2+}$ ions that run along the
crystallographic $a$ axis, and approximately parallel to the
$(a,c)$ plane. Spin correlations are ferromagnetic on the ladder
rung, but antiferromagnetic along the legs. The thus-formed
effective $S=1$ chain has a Haldane gap\cite{Haldane} of
$\Delta\approx 1.5$~meV,\cite{Manaka97,Masuda2006} seen at the 1D
AF zone-center $\mathbf{q}=(\frac{2n+1}{2},k,l)$, where $n$ is
integer. Due to weak interactions between ladders in the $(a,c)$
plane, the gap varies between $\Delta=1.2$~meV at $k=0$ (global
dispersion minimum) and $\Delta=1.8$~meV at $k=0.5$. Interactions
between ladders along the $b$ axis are negligible, due to entirely
non-magnetic sheets of organic molecules in-between. At low
temperatures \IPA\ goes through a field-induced AF ordering that
manifests itself in a lambda specific heat anomaly and  the
appearance of non-zero uniform magnetization.\cite{Manaka98} The
transition occurs at $H_{c,\mathbf{a}}=10.2$~T,
$H_{c,\mathbf{b}}=10.5$~T and $H_{c,\mathbf{c}}=10.4$~T, in fields
applied along the crystallographic $a^\ast$, $b^\ast$ and $c^\ast$
axes, respectively. This anisotropy effect is fully accounted for
by the anisotropy of the $g$-tensor for Cu$^{2+}$ in this system:
$g_{a}=2.05$ $g_{b}=2.22$ and
$g_{\bot}=2.11$.\cite{Manaka98,Manaka2007}

\section{Experimental}
In this work we summarize the results of several separate series
of neutron scattering experiments on deuterated single-crystal
\IPA\ samples. Measurements on the SPINS cold-neutron spectrometer
at the NIST Center for Neutron Research (NCNR) were performed on
an assembly of 20 single crystals of total mass 3g, mounted with
the crystallographic $b$ axis vertical, and the $(h,0,l)$
reciprocal space plane accessible for measurements (Setup I). The
mosaic of the assembly was somewhat irregular, with a FWHM of
4.5$^{\circ}$. Scattered neutrons were analyzed at 3.7~meV
 by a  pyrolitic graphite PG$(002)$
analyzer  in horizontal monochromatic-focusing mode, used in
combination with a BeO filter. A 120' radial collimator was
installed between sample and analyzer. A magnetic field of up to
11.5~T was applied along the $b$ axis. A dilution refrigerator
maintained sample temperature at 100~mK. In Setup II the sample
was mounted with the $c$ axis vertical (along the applied field)
and the $(h,k,0)$ plane open to scattering. An 80' collimation was
used in front of the sample, and only 5 out of 11 analyzer blades
were focused on the detector. Experiments in Setup III were
performed at the V2-FLEX 3-axis spectrometer at HMI. A smaller
crystal assembly of approximate mass 1~g was mounted with the $b$
axis vertical in a 14.5~T magnet. The data were collected at
70~mK, with no dedicated collimation devices in the neutron beam,
a PG filter after the sample, and a focusing PG (002) analyzer
tuned to 3.7~meV. A 15 Tesla magnet was used in Setup IV, where
the data were taken at $T<100$~mK on a 3g sample mounted with the
$c$-axis vertical on the IN-14 cold-neutron 3-axis spectrometer at
ILL. Sample mosaic was about 4$^{\circ}$ FWHM. Scattered neutrons
were analyzed at 2.9~meV by a focusing PG (002) analyzer with a Be
filter. A 60' beam collimator was used in front of the sample.

\begin{figure}
\includegraphics[width=8.7cm]{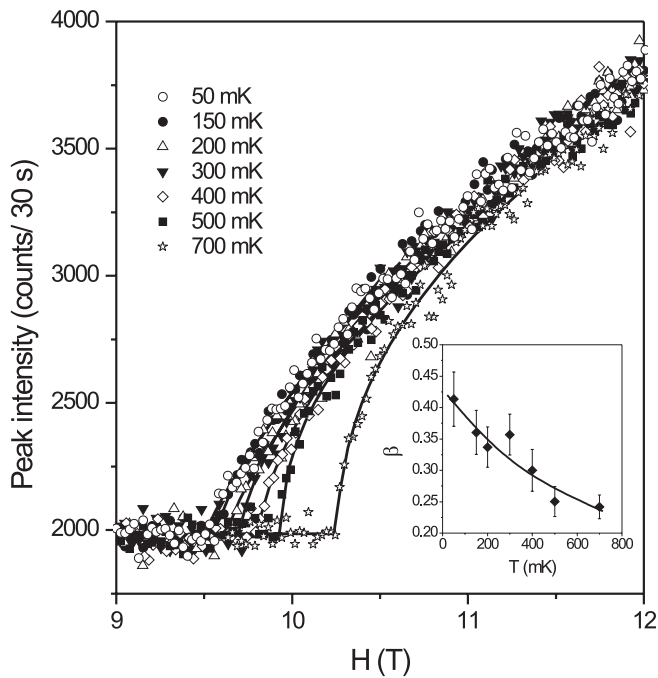}
\caption{Field dependence of the $(0.5,0,0)$ peak intensity
measured in at  different temperatures (symbols). The lines are
power-law fits to the data over a 1~T field range. Inset: The
value of $\beta$ fitted over a 1~T field range plotted as a
function of temperature. The line is a guide for the
eye.}\label{orderp}
\end{figure}

\begin{figure}
\includegraphics[width=8.7cm]{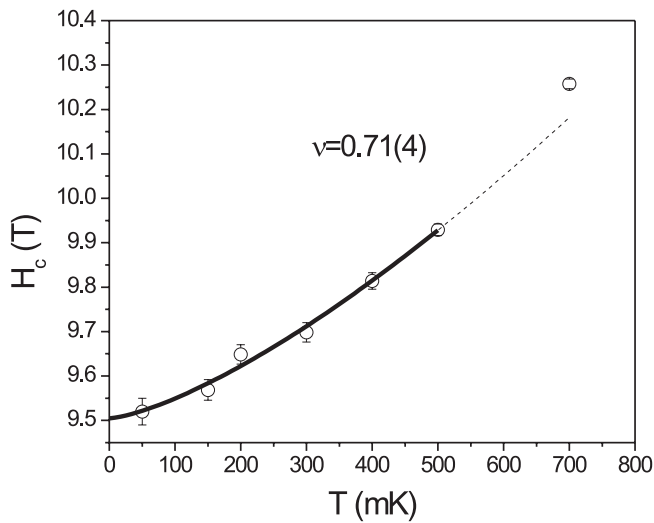}
\caption{Measured temperature dependence of the transition field
in \IPA\ (symbols). The line is a power-law fit to the data up to
$T=500$~mK.}\label{phase}
\end{figure}

In addition to these 3-axis data, the discussion below will
involve time-of-flight (TOF) spectra previously collected using
the Disk Chopper Spectrometer at NIST (Fig.~1 of
Ref.~\onlinecite{Garlea2007}). The sample in this experimental
Setup V was mounted with the $c$ axis vertical, parallel to the
applied field. Incident neutron energy was fixed at 6.7~meV.
Finally, diffraction data were taken on the D23 lifting counter
diffractometer, at ILL, using a monochromatic beam with $\lambda =
1.276~\AA$ (Setup VI). The measurements were carried out on a 3.2
x 2.3 x 8.5 mm$^3$ deuterated single crystal specimen. For this
experiment we used a 12~T vertical field superconducting magnet
with a dilution insert.

\section{Results}
\subsection{Field-induced AF long range order}
\subsubsection{Order parameter and phase diagram}
The main manifestation of the field-induced transition in \IPA\ is
the onset of commensurate transverse long-range AF order. The
high-field ordered phase is characterized by new Bragg reflections
of type $(\frac{2h+1}{2},k,l)$, with $h$, $k$ and
$l$-integers.\cite{Garlea2007}. The magnetic structure at $H=12$~T
applied along the $c$ direction was determined from 48 independent
Bragg intensities measured using Setup VI. The AF ordered moment
of $0.49(1)\mu_\mathrm{B}$ was found to be oriented almost
parallel to the crystallographic $b$ direction. As expected, the
two spins on each ladder rung are parallel to each other. The
field dependencies of the $(0.5,0,0)$ peak intensity measured at
several temperatures are shown in the main panel of
Fig.~\ref{orderp} (symbols).  The order parameter critical index
$\beta$ is defined as $m_\mathrm{AF}\propto (H-H_c)^\beta$, where
$m_\mathrm{AF}$ is the ordered AF staggered moment, and the
magnetic Bragg intensity scales as $I\propto m_\mathrm{AF}^2$. In
Ref.~\onlinecite{Garlea2007} we demonstrated that power law fits
over a progressively shrinking field range  at $T=50$~mK yield
$\beta$ close to $0.5$. In the present work, power-law fits were
performed over a fixed field range of 1~T above $H_c$ at each
temperature, and are plotted in solid lines in Fig.~\ref{orderp}.
The inset shows the thus measured temperature evolution of
$\beta$. This critical exponent rapidly decreases upon heating,
approaching the value $\beta=0.25$ at elevated temperatures. It
appears that with both the fitting range and temperature
approaching zero, $\beta$ will indeed approach the Mean Field
value of $0.5$, as expected for a BEC-type
transition.\cite{Giamarchi1999,Matsumoto2002}

Another characteristic of the phase transition is the critical
index $\nu$, defined as $T_c\propto (H-H_c)^\nu$. The measured
temperature dependence of the transition field is shown in
Fig.~\ref{phase}. A power law fit to this experimental phase
boundary gives $\nu=0.71(4)$. This value is close to $\nu=2/3$
expected for a 3D BEC of magnons,\cite{Giamarchi1999} but should
be regarded with some caution. Indeed, the density of data points
available at $T\rightarrow 0$ is barely sufficient for an accurate
estimate.

\subsection{Model dynamic structure factor} Before
discussing the results of inelastic measurements, we shall
introduce an analytical model magnetic neutron scattering cross
section for \IPA\ for use in the data analysis. A simple formula
that accurately describes the dispersion of the degenerate magnon
triplet at $H=0$ was discussed in Ref.~\onlinecite{Masuda2006}:
\begin{eqnarray}
 (\hbar \omega_\mathbf{q}^{(0)})^2&=&(\hbar\omega_0)^2\cos^2(\pi h) +  \Delta_0^2\sin^2(\pi
 h) +\nonumber\\
 &+& c_0^2\sin^2(2\pi h).\label{disp0}
\end{eqnarray}
The parameters were previously determined experimentally:
$\hbar\omega_0=4.08(9)$~meV, $\Delta_0=1.17(1)$~meV and
$c_0=2.15(9)$~meV. Equation \ref{disp0} follows the structural
periodicity of the spin ladder: wave vectors with $h$-integer and
$h$- half-integer are not, in general, equivalent. For AF spin
correlations $h=0.5$ is the 1D zone-center, while the zone
boundary is located at $h=0$ or $h=1$. Apart from the gap
$\Delta$, an important parameter is the velocity, that we defined
as
\begin{eqnarray}
v_0 & = & \frac{d \sqrt{(\hbar \omega_\mathbf{q}^{(0)})^2
-\Delta_0^2} }{ d (2\pi h)}\bigl{|}_{h=0.5}=
\nonumber\\
& =& \sqrt{c_0^2+[(\hbar \omega_0)^2 -\Delta_0^2]/4}=2.9
\mathrm{meV}.
\end{eqnarray}

In an external field the triplet spectrum is split due to Zeeman
effect. For $H<H_c$ the dispersion relation for the $i$-th magnon
branch with a field-dependent gap $\Delta^{(i)}$ can be written
as:
\begin{equation}
 \hbar\omega_\mathbf{q}^{(i)}=\hbar\omega_\mathbf{q}^{(0)}+\Delta_i-\Delta_0.\label{disp}
\end{equation}
Here we assume that the field only induces an overall energy
shift, without affecting the shape of the dispersion curve. Such
behavior is indeed consistent with experiment (see below), and
reflects the fact that the Zeeman term commutes with the
Heisenberg spin Hamiltonian.

The dynamic neutron cross section for each mode was written as:
\begin{equation}
 \frac{d^2 \sigma}{d \Omega d E'} \propto  |f(q)|^2
 \frac{A_i(\mathbf{q})}{\omega_\mathbf{q}^{(0)}}
 \delta(\omega-\omega_\mathbf{q}^{(i)}).\label{sqw}
\end{equation}
Here $f(q)$ is the magnetic form factor for Cu$^{2+}$ and
$A_i(\mathbf{q})$ is the overall intensity of mode $i$. The latter
implicitly includes the mode's structure factor $S_i(\mathbf{q})$,
the polarization factors for magnetic scattering of unpolarized
neutrons, as well as the appropriate components of the $g$-tensor.
The parameters of the model are the gap energies $\Delta_i$ and,
at each given wave vector, the intensities $A_i$ for each magnon
branch.

Equation~\ref{sqw} was also used to describe excitations in the
high field phase. However, due to the onset of AF long-range order
that doubles the structural period, at $H>H_\mathrm{c}$ $h=0$ and
$h=0.5$ become equivalent wave vectors, and the 1D zone-boundary
is shifted to $h=0.25$ or $h=0.75$. In this regime the dispersion
relation for the $i$-th branch was written as:
\begin{equation}
 (\hbar\omega_\mathbf{q}^{(i)})^2=\Delta_i^2+ c_i^2\sin^2(2\pi h).\label{disphf}
\end{equation}
The parameters  are the gap energies $\Delta_i$, velocities
$v_i\equiv c_i$, and, at any given wave vector, the intensities of
each mode.

\subsection{Spin hydrodynamics} Inelastic data collected near
the 1D AF zone-centers defined by half-integer values of $h$
provide information on the field dependence of the gap energies
and the polarizations of excitations.
\begin{figure}
\includegraphics[width=8.7cm]{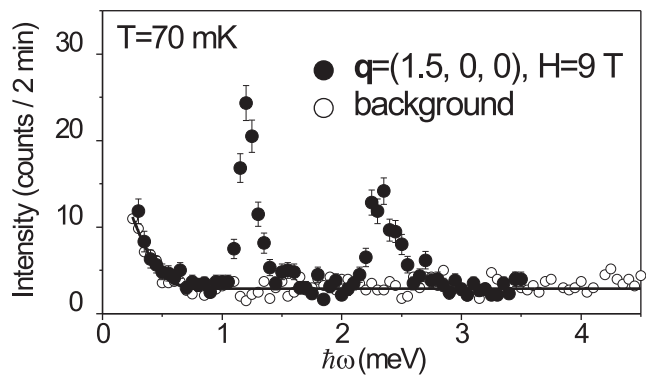}
\caption{Energy scan (raw data) collected in \IPA\ at $H=9$~T
using setup III, at $\mathbf{q}=(1.5,0,0)$ (solid symbols) and
background measured away from the 1D AF zone-center, at
$\mathbf{q}=(1.3,0,0)$ and $\mathbf{q}=(1.7,0,0)$ (open symbols).}
\label{rawdata3}
\end{figure}

\subsubsection{Gap energies} A representative
raw constant-$q$ scan measured at the 1D AF zone-center
$\mathbf{q}_1=(1.5,0,0)$ using Setup III at $H=9$~T is plotted in
Fig.~\ref{rawdata3} in solid symbols. The background was measured
at several fields as an average of scans at
$\mathbf{q}_1'=(1.3,0,0)$ and $\mathbf{q}_1''=(1.7,0,0)$. Since no
field dependence of the background was detected, it was further
averaged to include all experimental fields (open symbols in
Fig.~\ref{rawdata3}) and fit to a Gaussian centered at zero energy
transfer plus a constant (solid line). The resulting function was
then subtracted from all scans measured at the particular wave
vector transfer. A similar procedure was applied to constant-$q$
data at other wave vectors, or to those collected using most other
experimental setups.
\begin{figure}
\includegraphics[width=8.7cm]{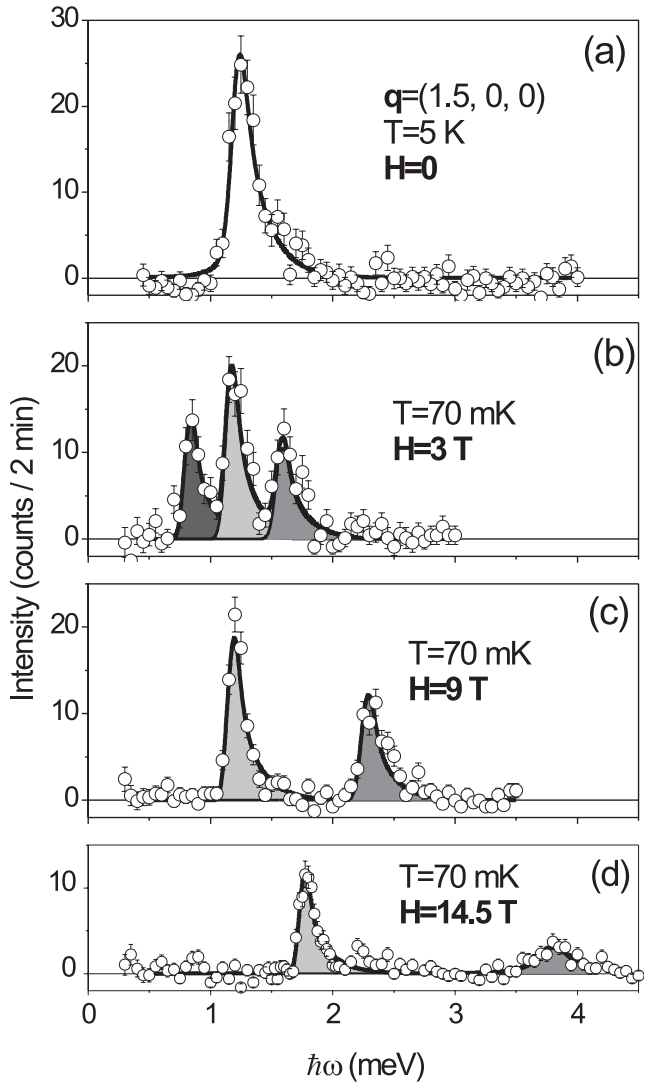}
\caption{Field dependence of background-subtracted inelastic
scattering in \IPA\ measured at the 1D AF zone-center
$\mathbf{q}_1=(1.5,0,0)$ using Setup III. Solid lines are model
cross section fits, as described in the text. Shaded areas are
partial contributions of three separate excitation branches.}
\label{exdata3}
\end{figure}

Background-subtracted scans collected at using Setup III at
various fields applied along the $\mathbf{b}$ axis are shown in
Fig.~\ref{exdata3}(a--d). As expected, the single peak seen at
$H=0$ splits into three components at higher fields. The gap in
the lower-energy mode extrapolates to zero at $H_c\approx 9.5$~T.
Close to and above the critical field only two of the three peaks
remain visible. In contrast with the previously studied
anisotropic Haldane gap compound NDMAP, the gap in the
lower-energy mode does not re-open above $H_c$.
\begin{figure}
\includegraphics[width=8.7cm]{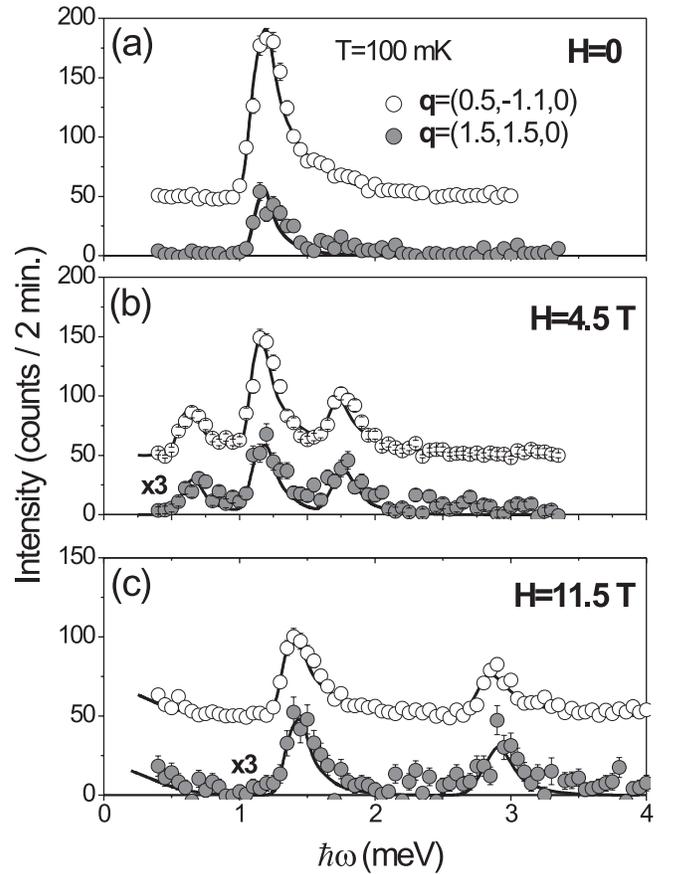}
\caption{Field dependence of background-subtracted inelastic
scattering in \IPA\ measured at two almost orthogonal wave vectors
$\mathbf{q}_2=(1.5,1.5,0)$ and $\mathbf{q}_3=(1.5,-1.1,0)$ using
Setup II. The reduced intensity at $\mathbf{q}_2$ is due to
neutron absorbtion in the sample. Solid lines are model cross
section fits, as described in the text. } \label{exdata2}
\end{figure}

\begin{figure}
\includegraphics[width=8.7cm]{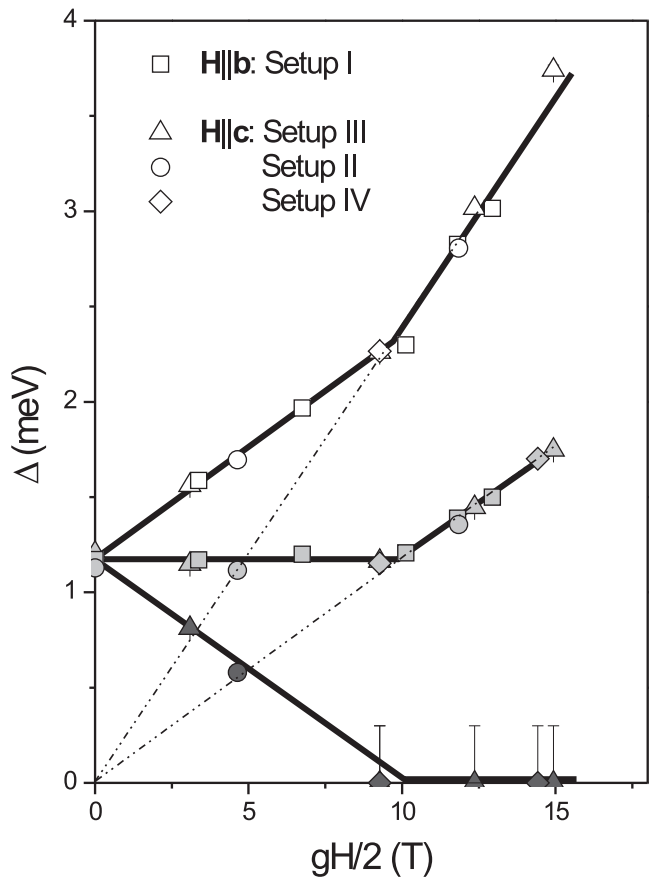}
\caption{Gap energies measured in \IPA\ at $T<100$~mK as a
function of reduced field for different experimental geometries
(symbols). Solid and dash-dot lines are guides for the eye. }
\label{gapvsh}
\end{figure}

Very similar data at equivalent wave vectors were collected using
Setup I for $H\|\mathbf{a}$, and for $H\|\mathbf{b}$ using Setup
II (Fig.~\ref{exdata2}) and Setup IV. In all cases, individual
scans were analyzed using the model cross section described in the
previous section. Eq.~\ref{sqw} was numerically convoluted with
the 4-dimensional $E-\mathbf{q}$ resolution function of the 3-axis
spectrometer, calculated in the Popovici approximation. At
$H>H_c$, Eq.~\ref{disphf} was utilized whenever the entire
dispersion relation of each mode was measured, and thus both
$\Delta_i$ and $c_i$ could be independently determined (see
below). In cases where the data collection was restricted to the
1D AF zone-center, Eq.~\ref{disp} was utilized instead, and only
the gap energies and intensities were varied. For $H<H_c$
Eq.~\ref{disp} was used in all cases. Typical fits are shown in
solid lines in Figs.~\ref{exdata3} and \ref{exdata2}. In
Fig.~\ref{exdata3} the shaded areas represent partial contribution
of each of the three members of the magnon triplet. The  field
dependencies of gap energies deduced from fits to the data
collected using all experimental conditions are summarized in
Fig.~\ref{gapvsh}. In order to compensate for the $g$-tensor
anisotropy, the abscissa in  actually shows the reduced field
value $gH/2$.

\subsubsection{Goldstone mode} 
\begin{figure}
\includegraphics[width=8.7cm]{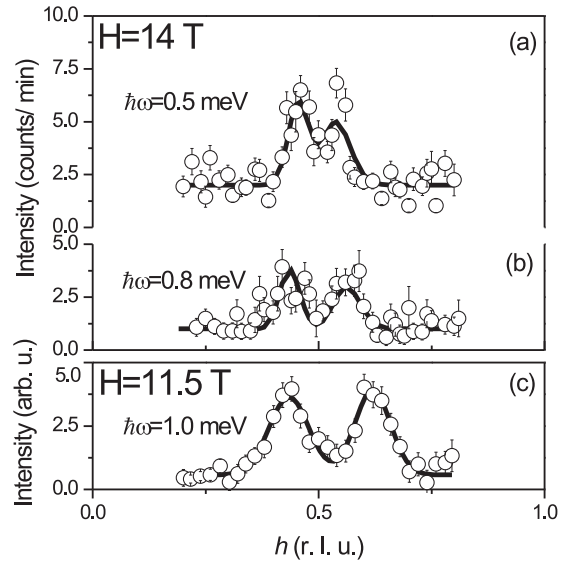}
\caption{$q$-scans measured in \IPA\ at low energies in the BEC
phase using Setup IV (a,b) and Setup V (c), showing the gapless
Goldstone mode. In (a) and (b) the solid lines are fits to the
data using a parameterized model cross section, as described in
the text. In (c) The solid lines are  empirical gaussian fits.}
\label{goldstone}
\end{figure}

As mentioned above, at $H>H_c$ the gap in the
lower mode in \IPA\ does not re-open. However, this does not mean
that the excitation itself disappears as, for instance, in the
$S=1$ dimerized chain system
NTENP.\cite{Hagiwara2005,Regnault2006} In \IPA\ the lower mode
persists above the critical field, but remains gapless, to within
experimental resolution. As discussed in
Ref.~\onlinecite{Garlea2007}, this excitation becomes the
Goldstone mode associated with the spontaneous breaking of
rotational symmetry due to magnetic ordering. It is best observed
in $q$-scans at low energy transfers, as shown in
Fig.~\ref{goldstone}. Here panels (a) and (b) display 3-axis data
collected using Setup IV. The solid lines are fits using
Eq.~\ref{disphf} with $\Delta=0$, and assuming a constant
background. Figure~\ref{goldstone}(c) shows a cut through the TOF
data of setup V taken between 0.8~meV and 1.2~meV energy transfer.
The solid lines are empyrical Gaussian fits. From this analysis we
obtain the spin wave velocities of the Goldstone mode
$v_\mathrm{G}=1.74$~meV and $v_\mathrm{G}=1.86$~meV, for
$H=11.5$~T and $H=14$~T applied along the $b$ axis, respectively.

\subsubsection{Intensities and polarizations of excitations} Since
magnetic interactions along the crystallographic $b$ axis are
totally  negligible, all wave vectors with the same values of $h$
and $l$ can be considered equivalent. In other words, the
$k$-dependence of magnetic scattering is fully decoupled from the
energy dependence. Polarization information can be deduced from a
comparison of constant-$q$ scans collected at different yet
equivalent wave vectors, thanks to neutrons being only scattered
by fluctuations of spin components perpendicular to the momentum
transfer.

\begin{figure}
\includegraphics[width=8.7cm]{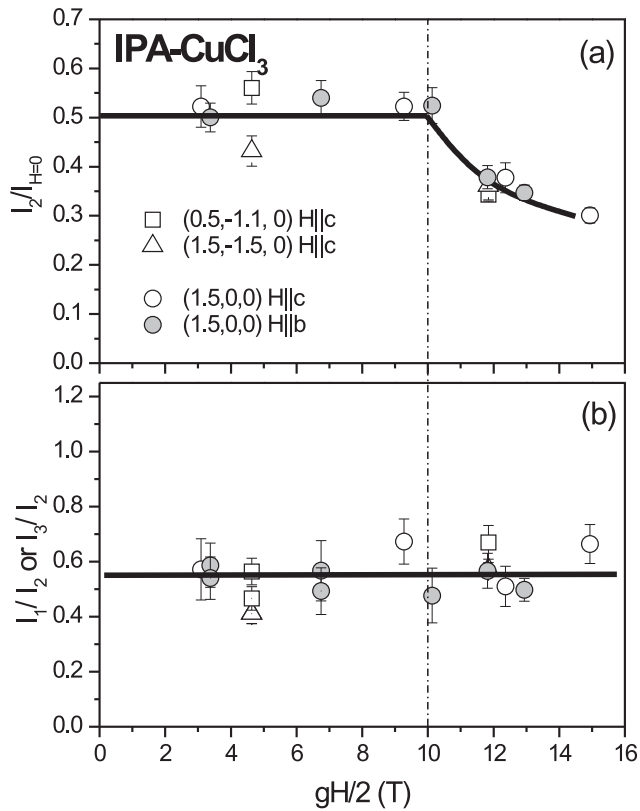}
\caption{(a) Intensity of the central magnon branch measured in
\IPA\ at $T<100$~mK as a function of reduced field for different
experimental geometries and scattering vectors (symbols). The
intensity is normalized to that of the single peak measured at
$H=0$. (b) Ratio of intensities of the upper and central (open
symbols) and lower and central (dot-centered symbols) magnon
branches measured in \IPA\ at $T<100$~mK as a function of reduced
field. In both panels the solid lines are guides for the eye. }
\label{intvsh}
\end{figure}

To take advantage of this selectivity, we compared scans collected
at three equivalent 1D zone-centers, separated by roughly
$45^\circ$ in reciprocal space. In Setups I and III the data were
taken at $\mathbf{q}_1=(1.5,0,0)$ (Fig.~\ref{exdata3}), to
complement the scans measured with Setup II at
$\mathbf{q}_2=(1.5,1.5,0)$ and $\mathbf{q}_3=(0.5,-1.1,0)$
(Fig.~\ref{exdata2}). These wave vectors form angles of
$\alpha_1=0$, $\alpha_3=46^\circ$, and $\alpha_3=-45^\circ$ with
the crystallographic $a^\ast$ axis, respectively. Experimentally,
for all geometries, the intensity of all gapped magnons remains
constant as long as $H<H_c$, but decreases above the transition
field. The intensity of the central magnon branch, deduced from
model cross section fits to constant-$q$ scans, is plotted against
field in Fig.~\ref{intvsh}(a). The data were normalized to the
intensity of the single inelastic peak seen at $H=0$.

 \begin{figure*}
\includegraphics[width=16 cm]{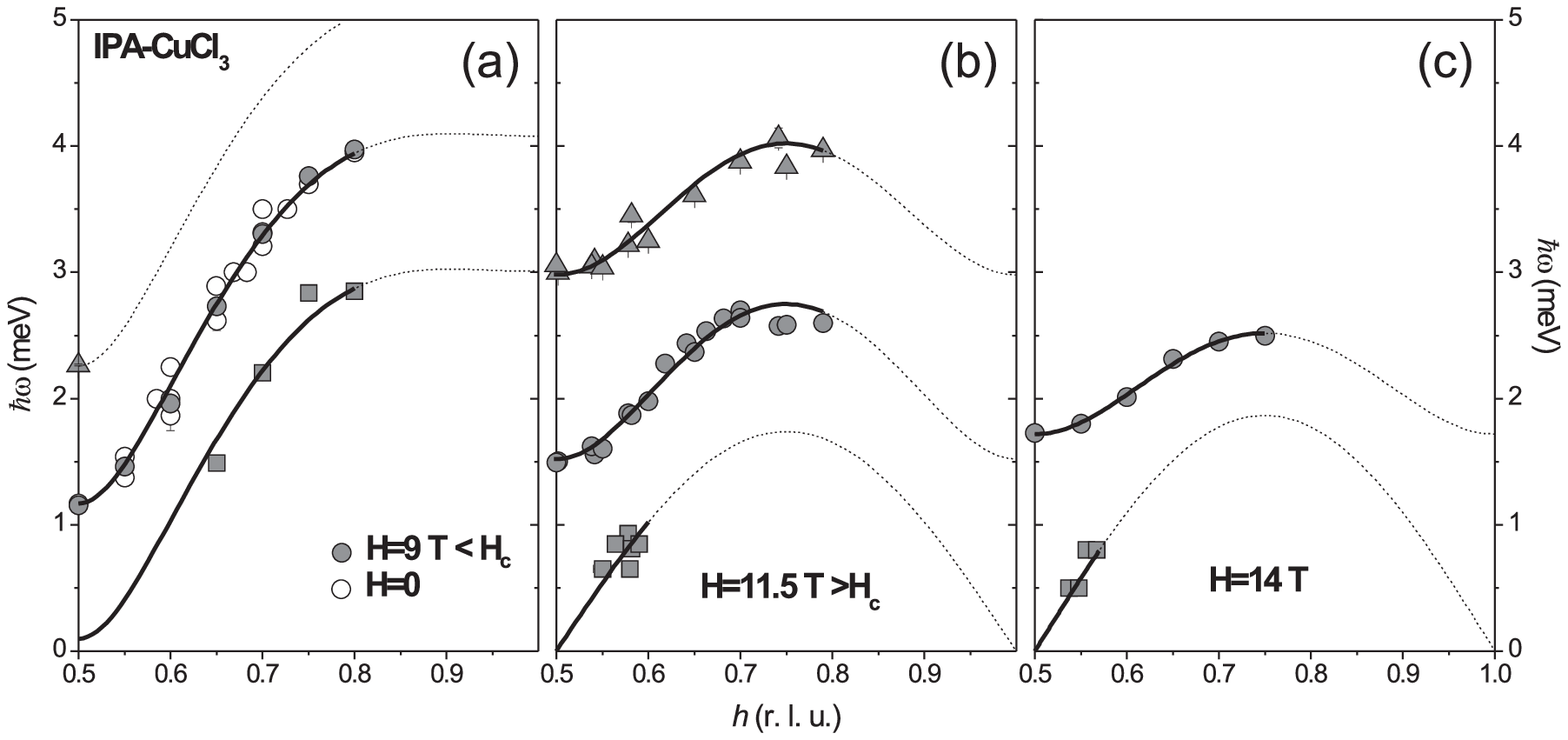}
\caption{Measured dispersion of magnons in \IPA\ below (a) and
above (b,c) the critical field. Symbols are fits to individual
constant-$q$ scans, and lines are semi-global fits at each field,
as described in the text. } \label{disp}
\end{figure*}

To within experimental statistics, the measured intensities of all
visible gapped magnons remain proportionate at all three wave
vectors in all fields. The corresponding intensity ratios are
plotted in Fig.~\ref{intvsh}(b). In all cases the central mode
remains roughly twice as strong the upper and lower ones, and half
as strong as the unresolved triplet peak at $H=0$. For $H<H_c$
this qualitative picture is consistent with expectations. Indeed,
the central mode is linearly polarized along the direction of
applied field ($z$-axis). It contributes equally to scattering at
all wave vectors in the $(x,y)$ scattering plane. The upper and
lower modes are expected to have circular polarization with
$S_z=\pm 1$. The contribution of each one is evenly split between
$S^{xx}(q_\|,\omega)$ and $S^{yy}(q_\|,\omega)$. If $x$ is chosen
along the scattering vector, only the $S^{yy}(q_\|,\omega)$
contribution is detected, resulting in a two-fold reduction of
intensity compared to linear polarization. A ratio of slightly
larger than $1/2$ is due to the anisotropy of the $g$-tensor in
\IPA. The magnetic neutron cross section for a given spin
polarization is proportional to the square of the corresponding
$g$-factor, which accounts for an enhancement of scattering in the
$(x,y)$ plane. A somewhat surprising experimental result is that
the intensity ratio of the upper and central modes remains
unchanged in the BEC phase. This behavior suggests that in
experimentally accessible field, even at $H>H_c$ where rotational
symmetry around the $z$ axis broken by long range ordering, the
intermixing of states with different polarization is not
particularly large.

\begin{figure}
\includegraphics[width=8.7cm]{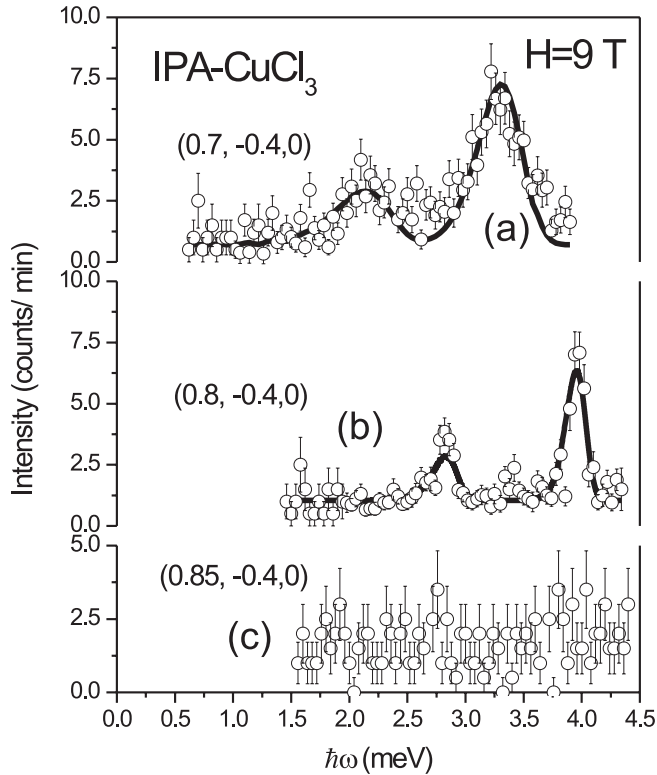}
\caption{Constant-$q$ scans measured in \IPA\ at $H=9$~T using
Setup IV (symbols). Solid lines are fits to the data as described
in the text. For the two visible branches of magnons the spectrum
is terminated at $h=\frac{1}{2}+h_c$, $h_c\approx 0.8$, as at
$H=0$ \protect\cite{Masuda2006}. } \label{termin9}
\end{figure}

\subsection{Spin dynamics away from the zone-center}
\subsubsection{Magnon dispersion curves}
The dispersion of magnons across the Brilloin zone was measured in
a series of constant-$q$ scans. The data were taken at $H=9$~T,
$H=11.5$~T and $H=14$~T using Setups IV and V. Due to geometric
limitations on accessible energy transfers at different wave
vectors, the dispersion relation of only the middle and lower
modes were measured in 3-axis experiments. These data were
analyzed using the model cross section described previously. A
semi-global fitting approach was implemented. Separate intensity
prefactors were used for each individual scan, but the same
dispersion relation was applied to all scans simultaneously at
each field. The results of these fits is shown in solid lines in
Fig.~\ref{disp}(a,c). Symbols represent fits to individual scans.
In Fig.~\ref{disp}(a) open symbols are data from
Ref.~\onlinecite{Masuda2006}, collected at $H=0$. For
Fig.~\ref{disp}(b), constant-$q_\|$ cutouts from the 2-dimensional
TOF data were fit to Gaussian profiles. The resulting dispersion
relation (symbols) was then analyzed using Eq.~\ref{disphf} (solid
lines). The key parameters of the fits are summarized in
Table~\ref{fitpars}. In all panels, a transition to dotted lines
indicate that either the area is outside the experimental scan
range or that no identifiable magnon peak was detected.

 \begin{table}
 \caption{\label{fitpars} Gap energies and spin wave velocities obtained
 from the measured magnon dispersion curves at three different fields. }
 \begin{ruledtabular}
 \begin{tabular}{l l l l }
 & $H=9$~T & $H=11.5$~T
 & $H=14$~T\\
 \hline\\
 $\Delta_1$ & 0.10(8)~meV      & 0 (fixed)& 0 (fixed)\\
 $\Delta_2$ & 1.17~meV (fixed) & 1.52(5)~meV  & 1.72(2)~meV\\
 $\Delta_3$ & 2.26(2)~meV      & 2.98(2)~meV  & -- \\
 \hline
 $v_1$      & 2.9~meV (fixed) & 1.74(3)~meV & 1.86(2)~meV\\
 $v_2$      & 2.9~meV (fixed) & 2.29(6)~meV & 1.84(3)~meV\\
 $v_3$      & 2.9~meV (fixed) & 2.67(4)~meV & --\\
 \end{tabular}
 \end{ruledtabular}
 \end{table}

\subsubsection{Spectrum termination}
One of the main findings of Ref.~\onlinecite{Masuda2006} was the
discovery of an abrupt truncation of the magnon branch in \IPA\ at
a critical wave vector $h=\frac{1}{2}+h_c$, $h_c=0.35$. This
phenomenon was attributed to 2-magnon decay, in analogy with
2-particle decay of rotons in liquid $^4$He. Similar physics was
observed in the quasi-2D system PHCC.\cite{Stone2006n} Further
theoretical studies\cite{Zhitomirsky2006,Kolezhuk2006} showed that
in one dimension the effect of 2-particle decay is particularly
dramatic, and always leads to a total removal of the
single-particle pole from the wave vector-energy domain of the
2-particle continuum. One of the goals of the present work was to
learn how the spectrum termination in \IPA\ is affected by an
external magnetic field, particularly above $H_c$.

\begin{figure}
\includegraphics[width=8.7cm]{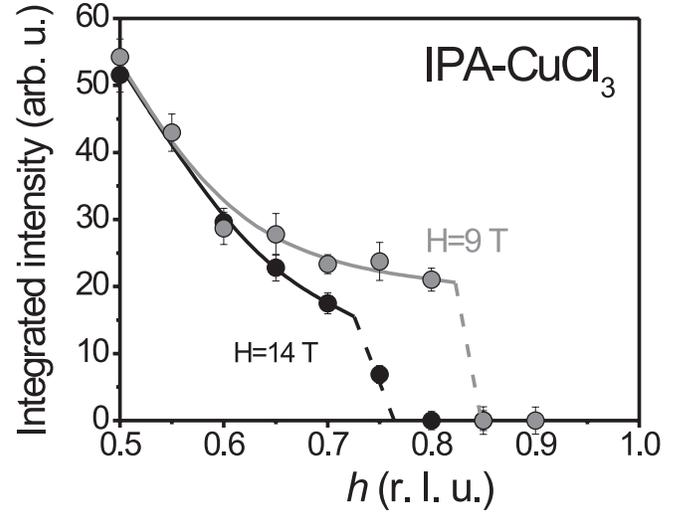}
\caption{Measured wave vector dependence of the intensity of the
middle magnon branch at $H=9$~T$<H_c$ (grey symbols) and
$H=14$~T$>H_c$ (solid symbols). Lines are guides for the eye.}
\label{termin}
\end{figure}

In Fig.~\ref{termin9} we show a series of raw constant-$q$ scans
collected using Setup IV near the maximum of the measured
dispersion, at $H=9$~T (just below the critical field). For
$h\lesssim \frac{1}{2}+h_c$, two peaks corresponding to the middle
and lower magnon branches are clearly visible
(Fig.~\ref{termin9}a,b). The 3rd magnon is outside the scan range.
The solid lines are fits using the model cross section given by
Eqs.~\ref{sqw} and \ref{disp}, convoluted with the spectrometer
resolution.   Both magnon peaks abruptly vanish at
$h>\frac{1}{2}+h_c$, as shown in Fig.~\ref{termin9}c. This
behavior is quantified by the plot of the fitted intensity of the
middle magnon branch, corrected for the magnetic form factor,
against $h$ in Fig.~\ref{termin} (grey symbols). We conclude that
at $H=9$~T the spectrum termination is exactly the same as in zero
field. Indeed, the trends visualized in Figure~\ref{termin9} are
very similar to those shown for  $H=0$ in Fig.~2(c,d) of
Ref.~\onlinecite{Masuda2006}.

\begin{figure}
\includegraphics[width=8.7cm]{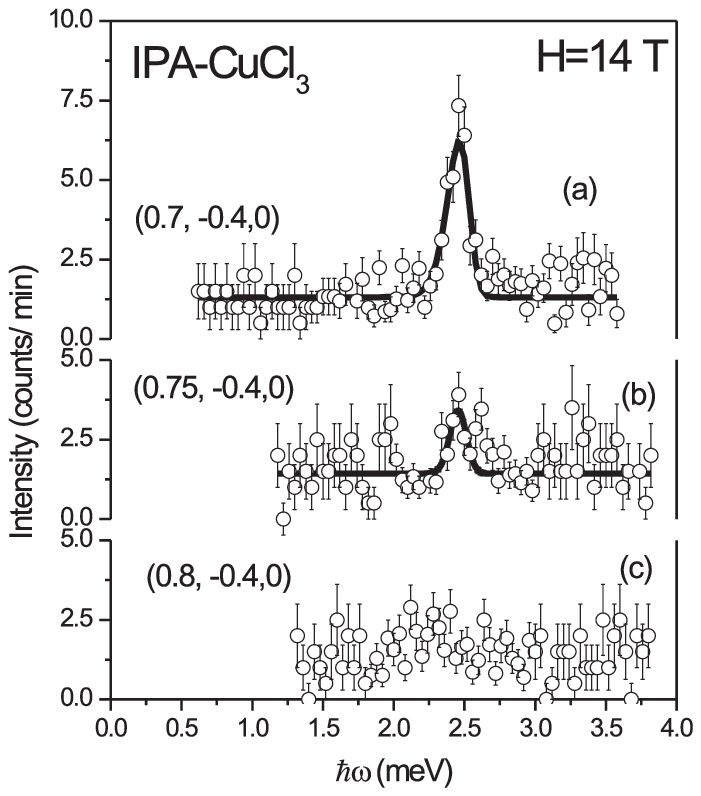}
\caption{Constant-$q$ scans measured in \IPA\ at $H=14$~T using
Setup IV (symbols). Solid lines are fits to the data as described
in the text. The magnon branch vanishes at
$h=\frac{1}{2}+h_c'$,$h_c'\approx 0.25$.} \label{termin14}
\end{figure}

The spectrum termination picture abruptly changes above the
critical field. Figure~\ref{termin14} shows the $h$-dependence of
constant-$q$ scans collected at $H=14$~T$>H_c$ using Setup IV.
Only the ``middle'' branch of magnons is visible, but does not
extend beyond $h\frac{1}{2}+h_c'$, $h_c'=0.25$, which in the high
field phase is the new 1D magnetic zone boundary. The
corresponding $h$-dependence of integrated magnon intensity is
shown in Fig.~\ref{termin} in solid circles.

In Fig.~\ref{termin14}a, note that at $h=0.7$ the observed peak is
much narrower than those seen at $H=9$~T (Fig.~\ref{termin9}a).
This, is entirely an effect of experimental wave vector
resolution. All observed width are in fact instrumental at all
fields. The narrow peak at $H=14$~T is due to that $h=0.7$ is
close to the new zone-boundary, so that the local slope of the
dispersion curve is shallow (Eq.~\ref{disphf}). In contrast, at
$H<H_c$, $h=0.75$ is still far from where the dispersion levels
out, and the local slope is steep (Eq.~\ref{disp}). A steep slope
leads to a broader peak, due to poor wave vector resolution, and a
relatively good energy resolution. Similar arguments explain why
at $H=9$~T at $h=0.8$ (Fig.~\ref{termin9}b) the magnons appear
narrower than and $h=0.7$.

\subsubsection{Bandwidth collapse}

A very important result reported in Ref.~\onlinecite{Garlea2007}
is the drastic reduction of the bandwidth of the middle (and,
probably, top) magnon branch at high fields. From Fig.~\ref{disp}a
we see that there is practically no change in the dispersion of
the middle magnon branch between $H=0$ and $H=9$~T, which means
that the bandwidth collapse occurs only at $H>H_c$. The bandwidth
continues to decrease in the entire accessible field range in the
high-field state, from 2.9~meV at $H=9$~T to as little at 0.8~meV
at $H=14$~T.

\begin{figure}
\includegraphics[width=8.7cm]{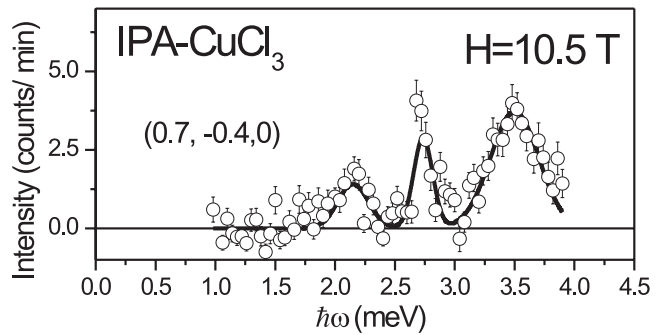}
\caption{A $\mathbf{q}=(0.7,-0.4,0)$ scan measured in \IPA\ at
$H=10.5$~T using Setup IV (symbols). Three peaks are observed, and
can be interpreted as a simultaneous observation of a 2-peak
feature as in Fig.~\protect\ref{termin9}a (upper and lower peaks)
and a single sharp peak as in Fig.~\protect\ref{termin14}a
(central peak).} \label{interm}
\end{figure}

A surprising finding is that between $H_c$ and $H=11.5$~T the
change of the spectrum of the central magnon is not continuous. In
fact, there appears to be a field range in which large-bandwidth
and low-bandwidth ``versions'' of the middle excitation branch
co-exist. This is illustrated by Fig.~\ref{interm} that shows a
scan taken at $h=0.7$ in an $H=10.5$~T external field. There
clearly are {\it three} peaks in this scan. The two broad peaks
can be associated with the two highly dispersive magnons seen
below $H_c$ at $H=9$~T (Fig.~\ref{termin9}a). At the same time,
the sharp peak resembles that corresponding to the middle mode at
higher fields (Fig.~\ref{termin14}a).

The two versions of the middle magnon branch apparently coexist
even at $H=11.5$~T, though the highly dispersive component is
considerably weakened. This can be inferred from the false color
plot of the TOF data in Fig.~1 of Ref.~\onlinecite{Garlea2007},
where a portion of the middle mode seems to continue all the way
to 4~meV as $h\rightarrow 0.8$, even as a stronger component
levels off at 2.7~meV. Due to poor experimental resolution, low
intensity, and a clear dominance of the narrow-bandwidth
component, it is not possible to clearly identify the more
dispersive component in constant-$q$ cut-outs at this field. At
$H=14$~T this excitation seems to have totally disappeared, and
only the narrow-bandwidth component is detected.

\section{Discussion}

The behavior of spin correlations in \IPA\ under field can be
summarized as follows.

1) At $H<H_c$ the system remains non-magnetic. All the way to the
critical field the three lower-energy magnon branches are exact
replicas of the degenerate magnon at $H=0$, but are shifted in
energy due to Zeeman effect, according to the spin projection that
they carry: $S_z=0,\pm 1$. This behavior is fully consistent with
the Zeeman term commuting with the Heisenberg Hamiltonian, and the
ground state being an $S=0$ singlet.

2) To within experimental accuracy, the critical properties of the
transition at $H_c$ and $T\rightarrow 0$ are those of BEC in a
dilute Bose gas. This is exactly what theory predicts in the case
of ideal axially symmetry.\cite{Giamarchi1999,Matsumoto2002} As
the softening magnons condense in the ground state at $H_c$, the
initial rotational $O(2)$ symmetry of the system is spontaneously
broken by the emerging AF long range order. This process is fully
equivalent to the formation of a conventional Bose-Einstein
condensate, where it is the $U(1)=O(2)$ gauge symmetry that is
being spontaneously violated by the macroscopic phase of the
condensate's wavefunction. In our case, at $T=0$ the density of
magnons is strictly zero for $H<H_c$, and vanishingly small just
above the transition. The phenomenon is then accurately described
in the limit of negligible quasiparticle interactions. Deviations
from this model become apparent with increasing temperature, due
to a thermal excitation of magnons and a finite magnon density at
the transition point. Experimentally, this thermal effect is
particularly strong for the order-parameter critical index $\beta$
that decreases rapidly with temperature.

3) The BEC nature for the phase transition is consistent with the
observed behavior of long-wavelength excitations near the AF point
at $H>H_c$. The lower of the three magnons becomes the Goldstone
mode, associated with the $O(2)$ symmetry breaking. To within
experimental resolution it remains gapless at all fields above
$H_c$. The intensity of this mode is decreased significantly
compared to that below the transition field.  At $H>H_c$ the gap
in the middle magnon branch begins to increase linearly.
Simultaneously, the rate of increase for the gap in the upper mode
doubles. The intensities of both the middle and upper modes are
suppressed with increasing field, while the velocities of all
three excitation branches progressively decrease.

This behavior is similar to that of other gapped quantum magnets
undergoing field-induced ordering transitions, such as the $S=1/2$
dimer compound TlCuCl$_3$, \cite{Ruegg2003} the $S=3/2$ material
Cs$_3$Cr$_2$Br$_9$,\cite{Grenier2004} or the Haldane spin chain
system NDMAP.\cite{Zheludev2003,Zheludev2004} The key difference,
of course, is that in those systems where magnetic anisotropy is
present, the transition is of an Ising, rather than BEC
universality class, and the lower mode re-acquires a gap in the
ordered phase.\cite{Affleck91} Field-theoretical descriptions of
gapped spin chains\cite{Affleck90-2,Affleck91}, as well as a
bond-operator treatment of the coupled-dimer
model,\cite{Matsumoto2002} attribute the changes of magnon gaps
and velocities to an admixture of the higher-energy triplet modes
to the ground state. This interpretation can be extended to our
case of coupled spin ladders. In particular, at $H>H_c$ rotational
symmetry is broken by the staggered moment, and $S_z$ ceases to be
a good quantum number for quasiparticles. The polarizations of all
magnon branches must become mixed. The Goldstone mode is thus
distinct from the $S_z=1$ magnon at $H<H_c$, just as the two gap
modes above the condensate can no longer be identified with
$S_z=0$ and $S_z=-1$ quasiparticles at low fields. For the
accessible range of magnetic fields, this polarization mixing in
\IPA\ can not be too strong though, as it was not directly
detected in our polarization studies. This is undoubtedly due to
the fact that even at $H=14$~T the ordered moment is still far
from its saturation value. On the other hand, the method used is,
admittedly, not the most accurate one: small shifts in
polarization will be easier to detect in future polarized neutron
experiments.

4) Short wave length spin dynamics of \IPA\ at $H>H_c$ is rather
complex. The Goldstone mode is not visible at higher energy
transfers, away from the zone-center, presumably due to its low
intensity. The middle magnon branch is actually replaced by {\it
two} excitations that merge at the zone-center, but are distinct
near the zone-boundary. One of these retains the steep dispersion
relation seen at $H<H_c$. This component progressively weakens
with increasing field and practically vanishes at $H\approx 12$~T.
Simultaneously, the 2nd component gains in intensity. It has a
much shallower dispersion, and entirely dominates at high fields.
The bandwidth of this component progressively decreases with
increasing $H$: the gap at the zone-center increase, and
zone-boundary energy goes down. The limited data that has been
accumulated on the 3rd (highest-energy) magnon branch suggest that
it's behavior is similar to that of the middle mode.

In our previous brief report the drastic restructuring of the
excitation spectrum above $H_c$ was attributed to a spontaneous
breaking of a discrete microscopic symmetry at the transition
point.\cite{Garlea2007} Unlike in a conventional BEC transition
that breaks a continuous $U(1)\equiv O(2)$ symmetry, the
transition in \IPA\ {\it in addition to  that} doubles the
magnetic structural period. The result is a folding of the
Brillouin zone and an opening of anticrossing gaps at the new zone
boundaries ($h=0.25$, $h=0.75$, and equivalent positions), where
each branch interacts with its own replica from an adjacent zone.
This, in turn, leads to an abrupt reduction of the zone-boundary
energy for the visible (lower) segments of the gapped magnons. The
additional breaking of translational symmetry is not an inherent
property of all magnon condensation transition, but specific to
\IPA\ and select other systems. For example, in the dimer material
TlCuCl$_3$ the periodicity of the magnetically ordered state is
the same as that of the underlying crystal
lattice.\cite{Ruegg2003} Correspondingly, the changes in the
excitation spectrum at the transition point\cite{Matsumoto2002}
are continuous and much less drastic than in \IPA.

The doubling of the ground state's translational periodicity is
also the primary cause for the shift of the critical wave vector
of spectrum termination at $H>H_c$. As discussed in detail in
Ref.~\onlinecite{Masuda2006}, at $H=0$ the magnon branch is
truncated at a point beyond which two-particle decay becomes
allowed by energy and momentum conservation. The particular shape
of the magnon dispersion curve in \IPA\ is such that the critical
wave vector is $h_c\approx 0.35$. Even though the magnon energies
shift at $0<H<H_c$, the termination point for each branch remains
unchanged in that regime. Due to conservation of spin projection,
the $S_z=0$ magnon, for example, can only decay into
quasiparticles with $S_z=-1$ and $S_z=1$. Since both the initial
and final states are eigenstates of the Zeeman term, none of the
matrix elements are affected by the applied field. Moreover, the
Zeeman shifts for the decay products are opposite and fully cancel
each other, so any energy and momentum conservation requirements
remain as at $H=0$. Similarly, below $H_c$ the field has no effect
on the decay of the $S_z=\pm 1$ particles.

This picture has to change qualitatively at $H>H_c$. The breaking
of rotational symmetry in the ground state removes the $S_z$
conservation requirement. It is easy to see that in this case
2-particle decay becomes possible at any wave vector beyond the
new zone-boundary $h=0.75$, so that $h_c'=0.25$. The final state
is a combination of the same type of magnon with a smaller wave
vector, and a Goldstone mode.

Certain important details regarding the short wavelength spin
dynamics of \IPA\ above the critical field are currently not
understood. In particular, we have no obvious explanation for the
apparent co-existence of wide- and narrow-bandwidth magnons just
above $H_c$.  Further theoretical studies of the problem are
clearly in order.

\section{Conclusion}
To date, most studies of field-induced ordering transitions in
quantum magnets were focused on long wave length spin
correlations. Indeed, these properties are directly relevant to
the phase transition itself. However, it appears that a lot is to
be learned and understood about short wavelength spin correlations
as well. This is particularly true when field-induced BEC of
magnons is accompanied by a breaking of discrete translational
symmetry, as in \IPA.

\acknowledgements Research at ORNL was funded by the United States
Department of Energy, Office of Basic Energy Sciences- Materials
Science, under Contract No. DE-AC05-00OR22725 with UT-Battelle,
LLC. T. M. was partially supported by the US - Japan Cooperative
Research Program on Neutron Scattering between the US DOE and
Japanese MEXT. The work at NIST is supported by the National
Science Foundation under Agreement Nos. DMR-9986442, -0086210, and
-0454672.


\begin{thebibliography}{26}
\expandafter\ifx\csname
natexlab\endcsname\relax\def\natexlab#1{#1}\fi
\expandafter\ifx\csname bibnamefont\endcsname\relax
  \def\bibnamefont#1{#1}\fi
\expandafter\ifx\csname bibfnamefont\endcsname\relax
  \def\bibfnamefont#1{#1}\fi
\expandafter\ifx\csname citenamefont\endcsname\relax
  \def\citenamefont#1{#1}\fi
\expandafter\ifx\csname url\endcsname\relax
  \def\url#1{\texttt{#1}}\fi
\expandafter\ifx\csname
urlprefix\endcsname\relax\def\urlprefix{URL }\fi
\providecommand{\bibinfo}[2]{#2}
\providecommand{\eprint}[2][]{\url{#2}}

\bibitem[{\citenamefont{Honda et~al.}(1997)\citenamefont{Honda, Katsumata,
  Katori, Yamada, Ohishi, Manabe, and Yamashita}}]{Honda97}
\bibinfo{author}{\bibfnamefont{Z.}~\bibnamefont{Honda}},
  \bibinfo{author}{\bibfnamefont{K.}~\bibnamefont{Katsumata}},
  \bibinfo{author}{\bibfnamefont{H.~A.} \bibnamefont{Katori}},
  \bibinfo{author}{\bibfnamefont{K.}~\bibnamefont{Yamada}},
  \bibinfo{author}{\bibfnamefont{T.}~\bibnamefont{Ohishi}},
  \bibinfo{author}{\bibfnamefont{T.}~\bibnamefont{Manabe}}, \bibnamefont{and}
  \bibinfo{author}{\bibfnamefont{M.}~\bibnamefont{Yamashita}},
  \bibinfo{journal}{J. Phys.: Condens. Matter} \textbf{\bibinfo{volume}{9}},
  \bibinfo{pages}{3487} (\bibinfo{year}{1997}).

\bibitem[{\citenamefont{Zheludev et~al.}(2003)\citenamefont{Zheludev, Honda,
  Broholm, Katsumata, Shapiro, Kolezhuk, Park, and Qiu}}]{Zheludev2003}
\bibinfo{author}{\bibfnamefont{A.}~\bibnamefont{Zheludev}},
  \bibinfo{author}{\bibfnamefont{Z.}~\bibnamefont{Honda}},
  \bibinfo{author}{\bibfnamefont{C.}~\bibnamefont{Broholm}},
  \bibinfo{author}{\bibfnamefont{K.}~\bibnamefont{Katsumata}},
  \bibinfo{author}{\bibfnamefont{S.~M.} \bibnamefont{Shapiro}},
  \bibinfo{author}{\bibfnamefont{A.}~\bibnamefont{Kolezhuk}},
  \bibinfo{author}{\bibfnamefont{S.}~\bibnamefont{Park}}, \bibnamefont{and}
  \bibinfo{author}{\bibfnamefont{Y.}~\bibnamefont{Qiu}},
  \bibinfo{journal}{Phys. Rev. B} \textbf{\bibinfo{volume}{68}},
  \bibinfo{pages}{134438} (\bibinfo{year}{2003}).

\bibitem[{\citenamefont{Zheludev et~al.}(2004)\citenamefont{Zheludev, Shapiro,
  Honda, Katsumata, Grenier, Ressouche, Regnault, Chen, Vorderwisch, Mikeska
  et~al.}}]{Zheludev2004}
\bibinfo{author}{\bibfnamefont{A.}~\bibnamefont{Zheludev}},
  \bibinfo{author}{\bibfnamefont{S.~M.} \bibnamefont{Shapiro}},
  \bibinfo{author}{\bibfnamefont{Z.}~\bibnamefont{Honda}},
  \bibinfo{author}{\bibfnamefont{K.}~\bibnamefont{Katsumata}},
  \bibinfo{author}{\bibfnamefont{B.}~\bibnamefont{Grenier}},
  \bibinfo{author}{\bibfnamefont{E.}~\bibnamefont{Ressouche}},
  \bibinfo{author}{\bibfnamefont{L.-P.} \bibnamefont{Regnault}},
  \bibinfo{author}{\bibfnamefont{Y.}~\bibnamefont{Chen}},
  \bibinfo{author}{\bibfnamefont{P.}~\bibnamefont{Vorderwisch}},
  \bibinfo{author}{\bibfnamefont{H.-J.} \bibnamefont{Mikeska}},
  \bibnamefont{et~al.}, \bibinfo{journal}{Phys. Rev. B}
  \textbf{\bibinfo{volume}{69}}, \bibinfo{pages}{054414}
  (\bibinfo{year}{2004}).

\bibitem[{\citenamefont{Narumi et~al.}(2001)\citenamefont{Narumi, Hagiwara,
  Kohno, and Kindo}}]{Narumi2001}
\bibinfo{author}{\bibfnamefont{Y.}~\bibnamefont{Narumi}},
  \bibinfo{author}{\bibfnamefont{M.}~\bibnamefont{Hagiwara}},
  \bibinfo{author}{\bibfnamefont{M.}~\bibnamefont{Kohno}}, \bibnamefont{and}
  \bibinfo{author}{\bibfnamefont{K.}~\bibnamefont{Kindo}},
  \bibinfo{journal}{Phys. Rev. Lett.} \textbf{\bibinfo{volume}{86}},
  \bibinfo{pages}{324} (\bibinfo{year}{2001}).

\bibitem[{\citenamefont{Hagiwara et~al.}(2005)\citenamefont{Hagiwara, Regnault,
  Zheludev, Stunault, Metoki, Suzuki, Suga, Kakurai, Koike, Vorderwisch
  et~al.}}]{Hagiwara2005}
\bibinfo{author}{\bibfnamefont{M.}~\bibnamefont{Hagiwara}},
  \bibinfo{author}{\bibfnamefont{L.~P.} \bibnamefont{Regnault}},
  \bibinfo{author}{\bibfnamefont{A.}~\bibnamefont{Zheludev}},
  \bibinfo{author}{\bibfnamefont{A.}~\bibnamefont{Stunault}},
  \bibinfo{author}{\bibfnamefont{N.}~\bibnamefont{Metoki}},
  \bibinfo{author}{\bibfnamefont{T.}~\bibnamefont{Suzuki}},
  \bibinfo{author}{\bibfnamefont{S.}~\bibnamefont{Suga}},
  \bibinfo{author}{\bibfnamefont{K.}~\bibnamefont{Kakurai}},
  \bibinfo{author}{\bibfnamefont{Y.}~\bibnamefont{Koike}},
  \bibinfo{author}{\bibfnamefont{P.}~\bibnamefont{Vorderwisch}},
  \bibnamefont{et~al.}, \bibinfo{journal}{Phys. Rev. Lett.}
  \textbf{\bibinfo{volume}{94}}, \bibinfo{pages}{177202}
  (\bibinfo{year}{2005}).


\bibitem{Regnault2006}  L.-P. Regnault, A. Zheludev, M. Hagiwara and A. Stunault,
Phys. Rev. B {\bf 73}, 174431 (2006).

\bibitem[{\citenamefont{Grenier et~al.}(2004)\citenamefont{Grenier, Inagaki,
  Regnault, Wildes, Asano, Ajiro, Lhotel, Paulsen, Ziman, and
  Boucher}}]{Grenier2004}
\bibinfo{author}{\bibfnamefont{B.}~\bibnamefont{Grenier}},
  \bibinfo{author}{\bibfnamefont{Y.}~\bibnamefont{Inagaki}},
  \bibinfo{author}{\bibfnamefont{L.}~\bibnamefont{Regnault}},
  \bibinfo{author}{\bibfnamefont{A.}~\bibnamefont{Wildes}},
  \bibinfo{author}{\bibfnamefont{T.}~\bibnamefont{Asano}},
  \bibinfo{author}{\bibfnamefont{Y.}~\bibnamefont{Ajiro}},
  \bibinfo{author}{\bibfnamefont{E.}~\bibnamefont{Lhotel}},
  \bibinfo{author}{\bibfnamefont{C.}~\bibnamefont{Paulsen}},
  \bibinfo{author}{\bibfnamefont{T.}~\bibnamefont{Ziman}}, \bibnamefont{and}
  \bibinfo{author}{\bibfnamefont{J.}~\bibnamefont{Boucher}},
  \bibinfo{journal}{Phys. Rev. Lett.} \textbf{\bibinfo{volume}{92}},
  \bibinfo{pages}{177202} (\bibinfo{year}{2004}).

\bibitem[{\citenamefont{Ziman et~al.}(2005)\citenamefont{Ziman, Boucher,
  Inagaki, and Ajiro}}]{Ziman2005}
\bibinfo{author}{\bibfnamefont{T.}~\bibnamefont{Ziman}},
  \bibinfo{author}{\bibfnamefont{J.~P.} \bibnamefont{Boucher}},
  \bibinfo{author}{\bibfnamefont{Y.}~\bibnamefont{Inagaki}}, \bibnamefont{and}
  \bibinfo{author}{\bibfnamefont{Y.}~\bibnamefont{Ajiro}}, \bibinfo{journal}{J.
  Phys. Soc. Jpn.} \textbf{\bibinfo{volume}{74 Suppl.}}, \bibinfo{pages}{119}
  (\bibinfo{year}{2005}).

\bibitem[{\citenamefont{Affleck}(1991)}]{Affleck91}
\bibinfo{author}{\bibfnamefont{I.}~\bibnamefont{Affleck}},
  \bibinfo{journal}{Phys. Rev. B} \textbf{\bibinfo{volume}{43}},
  \bibinfo{pages}{3215} (\bibinfo{year}{1991}).

\bibitem[{\citenamefont{Ruegg et~al.}(2003)\citenamefont{Ruegg, Cavadini,
  Furrer, Gudel, Kramer, Mutka, Wildes, Habicht, and Vorderwisch}}]{Ruegg2003}
\bibinfo{author}{\bibfnamefont{C.}~\bibnamefont{Ruegg}},
  \bibinfo{author}{\bibfnamefont{N.}~\bibnamefont{Cavadini}},
  \bibinfo{author}{\bibfnamefont{A.}~\bibnamefont{Furrer}},
  \bibinfo{author}{\bibfnamefont{H.-U.} \bibnamefont{Gudel}},
  \bibinfo{author}{\bibfnamefont{K.}~\bibnamefont{Kramer}},
  \bibinfo{author}{\bibfnamefont{H.}~\bibnamefont{Mutka}},
  \bibinfo{author}{\bibfnamefont{A.}~\bibnamefont{Wildes}},
  \bibinfo{author}{\bibfnamefont{K.}~\bibnamefont{Habicht}}, \bibnamefont{and}
  \bibinfo{author}{\bibfnamefont{P.}~\bibnamefont{Vorderwisch}},
  \bibinfo{journal}{Nature} \textbf{\bibinfo{volume}{423}}, \bibinfo{pages}{62}
  (\bibinfo{year}{2003}).

\bibitem[{\citenamefont{Stone et~al.}(2006)\citenamefont{Stone, Broholm, Reich,
  Tchernyshev, Vorderwisch, and Harrison}}]{Stone2006}
\bibinfo{author}{\bibfnamefont{M.~B.} \bibnamefont{Stone}},
  \bibinfo{author}{\bibfnamefont{C.}~\bibnamefont{Broholm}},
  \bibinfo{author}{\bibfnamefont{D.~H.} \bibnamefont{Reich}},
  \bibinfo{author}{\bibfnamefont{O.}~\bibnamefont{Tchernyshev}},
  \bibinfo{author}{\bibfnamefont{P.}~\bibnamefont{Vorderwisch}},
  \bibnamefont{and} \bibinfo{author}{\bibfnamefont{N.}~\bibnamefont{Harrison}},
  \bibinfo{journal}{Phys. Rev. Lett.} \textbf{\bibinfo{volume}{96}},
  \bibinfo{pages}{257203} (\bibinfo{year}{2006}).

\bibitem[{\citenamefont{Sebastian et~al.}(2005)\citenamefont{Sebastian, Sharma,
  Jaime, Harrison, Correa, Balicas, Kawashima, Batista, and
  Fisher}}]{Sebastian2005}
\bibinfo{author}{\bibfnamefont{S.~E.} \bibnamefont{Sebastian}},
  \bibinfo{author}{\bibfnamefont{P.~A.} \bibnamefont{Sharma}},
  \bibinfo{author}{\bibfnamefont{M.}~\bibnamefont{Jaime}},
  \bibinfo{author}{\bibfnamefont{N.}~\bibnamefont{Harrison}},
  \bibinfo{author}{\bibfnamefont{V.}~\bibnamefont{Correa}},
  \bibinfo{author}{\bibfnamefont{L.}~\bibnamefont{Balicas}},
  \bibinfo{author}{\bibfnamefont{N.}~\bibnamefont{Kawashima}},
  \bibinfo{author}{\bibfnamefont{C.~D.} \bibnamefont{Batista}},
  \bibnamefont{and} \bibinfo{author}{\bibfnamefont{I.~R.}
  \bibnamefont{Fisher}}, \bibinfo{journal}{Phys. Rev. B}
  \textbf{\bibinfo{volume}{72}}, \bibinfo{pages}{100404(R)}
  (\bibinfo{year}{2005}).

\bibitem[{\citenamefont{Zapf et~al.}(2006)\citenamefont{Zapf, Zocco, Hansen,
  Jaime, Batista, Kenzelmann, Niedermayer, Lacerda, and
  Paduan-Filho}}]{Zapf2006}
\bibinfo{author}{\bibfnamefont{V.~S.} \bibnamefont{Zapf}},
  \bibinfo{author}{\bibfnamefont{D.}~\bibnamefont{Zocco}},
  \bibinfo{author}{\bibfnamefont{B.~R.} \bibnamefont{Hansen}},
  \bibinfo{author}{\bibfnamefont{M.}~\bibnamefont{Jaime}},
  \bibinfo{author}{\bibfnamefont{C.~D.} \bibnamefont{Batista}},
  \bibinfo{author}{\bibfnamefont{M.}~\bibnamefont{Kenzelmann}},
  \bibinfo{author}{\bibfnamefont{C.}~\bibnamefont{Niedermayer}},
  \bibinfo{author}{\bibfnamefont{A.}~\bibnamefont{Lacerda}}, \bibnamefont{and}
  \bibinfo{author}{\bibfnamefont{A.}~\bibnamefont{Paduan-Filho}},
  \bibinfo{journal}{Phys. Rev. Lett.} \textbf{\bibinfo{volume}{96}},
  \bibinfo{pages}{077204} (\bibinfo{year}{2006}).

\bibitem[{\citenamefont{Giamarchi and Tsvelik}(1999)}]{Giamarchi1999}
\bibinfo{author}{\bibfnamefont{T.}~\bibnamefont{Giamarchi}} \bibnamefont{and}
  \bibinfo{author}{\bibfnamefont{A.~M.} \bibnamefont{Tsvelik}},
  \bibinfo{journal}{Phys. Rev. B} \textbf{\bibinfo{volume}{59}},
  \bibinfo{pages}{11398} (\bibinfo{year}{1999}).

\bibitem[{\citenamefont{Mills}(2007)}]{Mills2007}
\bibinfo{author}{\bibfnamefont{D.~L.} \bibnamefont{Mills}},
  \bibinfo{journal}{Phys. Rev. Lett.} \textbf{\bibinfo{volume}{98}},
  \bibinfo{pages}{039701} (\bibinfo{year}{2007}).

\bibitem[{\citenamefont{Sirker et~al.}(2005)\citenamefont{Sirker, Weiße, and
  Sushkov}}]{Sirker2005}
\bibinfo{author}{\bibfnamefont{J.}~\bibnamefont{Sirker}},
  \bibinfo{author}{\bibfnamefont{A.}~\bibnamefont{Weiße}}, \bibnamefont{and}
  \bibinfo{author}{\bibfnamefont{O.~P.} \bibnamefont{Sushkov}},
  \bibinfo{journal}{J. Phys. Soc. Jpn.} \textbf{\bibinfo{volume}{74 Suppl.}},
  \bibinfo{pages}{129} (\bibinfo{year}{2005}).

\bibitem[{\citenamefont{Johannsen et~al.}(2005)\citenamefont{Johannsen,
  Vasiliev, Oosawa, Tanaka, , and Lorenz}}]{Johannsen2005}
\bibinfo{author}{\bibfnamefont{N.}~\bibnamefont{Johannsen}},
  \bibinfo{author}{\bibfnamefont{A.}~\bibnamefont{Vasiliev}},
  \bibinfo{author}{\bibfnamefont{A.}~\bibnamefont{Oosawa}},
  \bibinfo{author}{\bibfnamefont{H.}~\bibnamefont{Tanaka}}, , \bibnamefont{and}
  \bibinfo{author}{\bibfnamefont{T.}~\bibnamefont{Lorenz}},
  \bibinfo{journal}{Phys. Rev. Lett.} \textbf{\bibinfo{volume}{95}},
  \bibinfo{pages}{017205} (\bibinfo{year}{2005}).

\bibitem[{\citenamefont{Garlea et~al.}(2007)\citenamefont{Garlea, Zheludev,
  Masuda, Manaka, Regnault, Ressouche, Grenier, Chung, Qiu, Habicht
  et~al.}}]{Garlea2007}
\bibinfo{author}{\bibfnamefont{V.~O.} \bibnamefont{Garlea}},
  \bibinfo{author}{\bibfnamefont{A.}~\bibnamefont{Zheludev}},
  \bibinfo{author}{\bibfnamefont{T.}~\bibnamefont{Masuda}},
  \bibinfo{author}{\bibfnamefont{H.}~\bibnamefont{Manaka}},
  \bibinfo{author}{\bibfnamefont{L.-P.} \bibnamefont{Regnault}},
  \bibinfo{author}{\bibfnamefont{E.}~\bibnamefont{Ressouche}},
  \bibinfo{author}{\bibfnamefont{B.}~\bibnamefont{Grenier}},
  \bibinfo{author}{\bibfnamefont{J.-H.} \bibnamefont{Chung}},
  \bibinfo{author}{\bibfnamefont{Y.}~\bibnamefont{Qiu}},
  \bibinfo{author}{\bibfnamefont{K.}~\bibnamefont{Habicht}},
  \bibnamefont{et~al.}, \bibinfo{journal}{Phys. Rev. Lett.}
  \textbf{\bibinfo{volume}{98}}, \bibinfo{pages}{167202}
  (\bibinfo{year}{2007}).

\bibitem[{\citenamefont{Masuda et~al.}(2006)\citenamefont{Masuda, Zheludev,
  Manaka, Regnault, Chung, and Qiu}}]{Masuda2006}
\bibinfo{author}{\bibfnamefont{T.}~\bibnamefont{Masuda}},
  \bibinfo{author}{\bibfnamefont{A.}~\bibnamefont{Zheludev}},
  \bibinfo{author}{\bibfnamefont{H.}~\bibnamefont{Manaka}},
  \bibinfo{author}{\bibfnamefont{L.-P.} \bibnamefont{Regnault}},
  \bibinfo{author}{\bibfnamefont{J.-H.} \bibnamefont{Chung}}, \bibnamefont{and}
  \bibinfo{author}{\bibfnamefont{Y.}~\bibnamefont{Qiu}},
  \bibinfo{journal}{Phys. Rev. Lett.} \textbf{\bibinfo{volume}{96}},
  \bibinfo{pages}{047210} (\bibinfo{year}{2006}).

\bibitem[{\citenamefont{Manaka et~al.}(1997)\citenamefont{Manaka, Yamada, and
  Yamaguchi}}]{Manaka97}
\bibinfo{author}{\bibfnamefont{H.}~\bibnamefont{Manaka}},
  \bibinfo{author}{\bibfnamefont{I.}~\bibnamefont{Yamada}}, \bibnamefont{and}
  \bibinfo{author}{\bibfnamefont{K.}~\bibnamefont{Yamaguchi}},
  \bibinfo{journal}{J. Phys. Soc. Jpn.} \textbf{\bibinfo{volume}{66}},
  \bibinfo{pages}{564} (\bibinfo{year}{1997}).

\bibitem[{Hal()}]{Haldane}
\bibinfo{note}{F. D. M. Haldane, Phys. Lett. {\bf 93A}, 464 (1983); Phys. Rev.
  Lett. {\bf 50}, 1153(1983).}

\bibitem[{\citenamefont{Manaka et~al.}(1998)\citenamefont{Manaka, Yamada,
  Honda, Katori, and Katsumata}}]{Manaka98}
\bibinfo{author}{\bibfnamefont{H.}~\bibnamefont{Manaka}},
  \bibinfo{author}{\bibfnamefont{I.}~\bibnamefont{Yamada}},
  \bibinfo{author}{\bibfnamefont{Z.}~\bibnamefont{Honda}},
  \bibinfo{author}{\bibfnamefont{H.~A.} \bibnamefont{Katori}},
  \bibnamefont{and}
  \bibinfo{author}{\bibfnamefont{K.}~\bibnamefont{Katsumata}},
  \bibinfo{journal}{J. Phys. Soc. Jpn.} \textbf{\bibinfo{volume}{67}},
  \bibinfo{pages}{3913} (\bibinfo{year}{1998}).


\bibitem{Manaka2007} H. Manaka, K. Masamoto, and S. Maehata, J. Phys. Soc. Jpn. {\bf 76},
023002 (2007).




\bibitem[{\citenamefont{Matsumoto et~al.}(2002)\citenamefont{Matsumoto,
  Normand, Rice, and Sigrist}}]{Matsumoto2002}
\bibinfo{author}{\bibfnamefont{M.}~\bibnamefont{Matsumoto}},
  \bibinfo{author}{\bibfnamefont{B.}~\bibnamefont{Normand}},
  \bibinfo{author}{\bibfnamefont{T.~M.} \bibnamefont{Rice}}, \bibnamefont{and}
  \bibinfo{author}{\bibfnamefont{M.}~\bibnamefont{Sigrist}},
  \bibinfo{journal}{Phys. Rev. Lett.} \textbf{\bibinfo{volume}{89}},
  \bibinfo{pages}{077203} (\bibinfo{year}{2002}).

\bibitem[{\citenamefont{M.~B.~Stone}(2006)}]{Stone2006n}
\bibinfo{author}{\bibfnamefont{T.~H. C. L. B. D. H.~R.}
  \bibnamefont{M.~B.~Stone}, \bibfnamefont{I.~A.~Zaliznyak}},
  \bibinfo{journal}{Nature} \textbf{\bibinfo{volume}{440}},
  \bibinfo{pages}{190} (\bibinfo{year}{2006}).

\bibitem[{\citenamefont{Zhitomirsky}(2006)}]{Zhitomirsky2006}
\bibinfo{author}{\bibfnamefont{M.}~\bibnamefont{Zhitomirsky}},
  \bibinfo{journal}{Phys. Rev. B} \textbf{\bibinfo{volume}{73}},
  \bibinfo{pages}{100404} (\bibinfo{year}{2006}).

\bibitem[{\citenamefont{Kolezhuk and Sachdev}(2006)}]{Kolezhuk2006}
\bibinfo{author}{\bibfnamefont{A.}~\bibnamefont{Kolezhuk}} \bibnamefont{and}
  \bibinfo{author}{\bibfnamefont{S.}~\bibnamefont{Sachdev}},
  \bibinfo{journal}{Phys. Rev. Lett.} \textbf{\bibinfo{volume}{96}},
  \bibinfo{pages}{087203} (\bibinfo{year}{2006}).

\bibitem[{\citenamefont{Affleck}(1990)}]{Affleck90-2}
\bibinfo{author}{\bibfnamefont{I.}~\bibnamefont{Affleck}},
  \bibinfo{journal}{Phys. Rev. B} \textbf{\bibinfo{volume}{41}},
  \bibinfo{pages}{6697} (\bibinfo{year}{1990}).

\end{thebibliography}
\end{document}